\documentclass[3p,twocolumn]{elsarticle}

\usepackage{amsmath,amssymb}
\usepackage{graphicx}
\usepackage{lineno}
\usepackage{geometry}

\usepackage{hyperref}
\hypersetup{hidelinks}
\journal{Science Bulletin}

\makeatletter
\renewcommand{\figurename}{Fig.}
\renewcommand{\fnum@figure}{\textbf{\figurename~\thefigure.}}
\long\def\@makecaption#1#2{%
  \vskip\abovecaptionskip
  \sbox\@tempboxa{#1 #2}
  \ifdim \wd\@tempboxa >\hsize
    #1 #2\par
  \else
    \global \@minipagefalse
    \hb@xt@\hsize{\hfil\box\@tempboxa\hfil}%
  \fi
  \vskip\belowcaptionskip}
\def\ps@pprintTitle{%
 \let\@oddhead\@empty
 \let\@evenhead\@empty
 \def\@oddfoot{\reset@font\hfil\thepage\hfil}
 \let\@evenfoot\@oddfoot
}
\makeatother

\begin{document}

\begin{frontmatter}

\title{A Diamond Quantum Oscilloscope}

\author[a,b,c,d]{Yang Dong}
\author[a,b,c]{Yu Zheng}
\author[a,b,c]{Wang Jiang}
\author[a,b,c]{Zhi-Wei Liu}
\author[a,b,c]{Shao-Chun Zhang}
\author[a,b,c,e]{Xiang-Dong Chen}
\author[a,b,c,e]{Guang-Can Guo}
\author[a,b,c,e]{Fang-Wen Sun\corref{cor1}}
\cortext[cor1]{Corresponding author: fwsun@ustc.edu.cn}

\address[a]{Laboratory of Quantum Information, University of Science and Technology of China, Hefei 230026, P.R. China}
\address[b]{Anhui Province Key Laboratory of Quantum Network, University of Science and Technology of China, Hefei 230026, P.R. China}
\address[c]{CAS Center for Excellence in Quantum Information and Quantum Physics, University of Science and Technology of China, Hefei 230026, P.R. China}
\address[d]{School of Physics, Harbin Institute of Technology, Harbin 150006, P.R. China}
\address[e]{Hefei National Laboratory, University of Science and Technology of China, Hefei 230088, P.R. China}

\begin{abstract}
Quantum-enhanced metrology has achieved remarkable precision in parameter estimation, but extending these advantages to continuous time-dependent fields (waveform sensing) remains a formidable challenge due to the infinite-dimensional nature of the signal. Here, we report the first experimental realization of a room-temperature solid-state quantum oscilloscope based on a sequential quantum difference amplification strategy. Within a zero-order-hold reconstruction framework, our method completely bypasses the need for massive entanglement. Crucially, we demonstrate that the overall waveform reconstruction error unequivocally follows the theoretically predicted Heisenberg scaling of \(O(N^{-1/2})\), decisively surpassing the classical \(O(N^{-1/3})\) limit. This transformative precision requires approximately 30-fold fewer quantum resources than conventional methods to achieve the same waveform error. Furthermore, we demonstrate the reconstruction of a multicomponent waveform and distinctly resolve weak high-frequency components that are entirely drowned in the shot noise of classical sampling. These results transition quantum metrology from parameter estimation into a practical instrument, establishing a genuine quantum advantage in continuous waveform reconstruction and paving the way for critical applications such as nanoscale nuclear magnetic resonance imaging, where detecting extremely weak signals is crucial for resolving molecular structures.
\end{abstract}

\end{frontmatter}

\begin{figure}[t]
\centering
\includegraphics[width=0.5\textwidth]{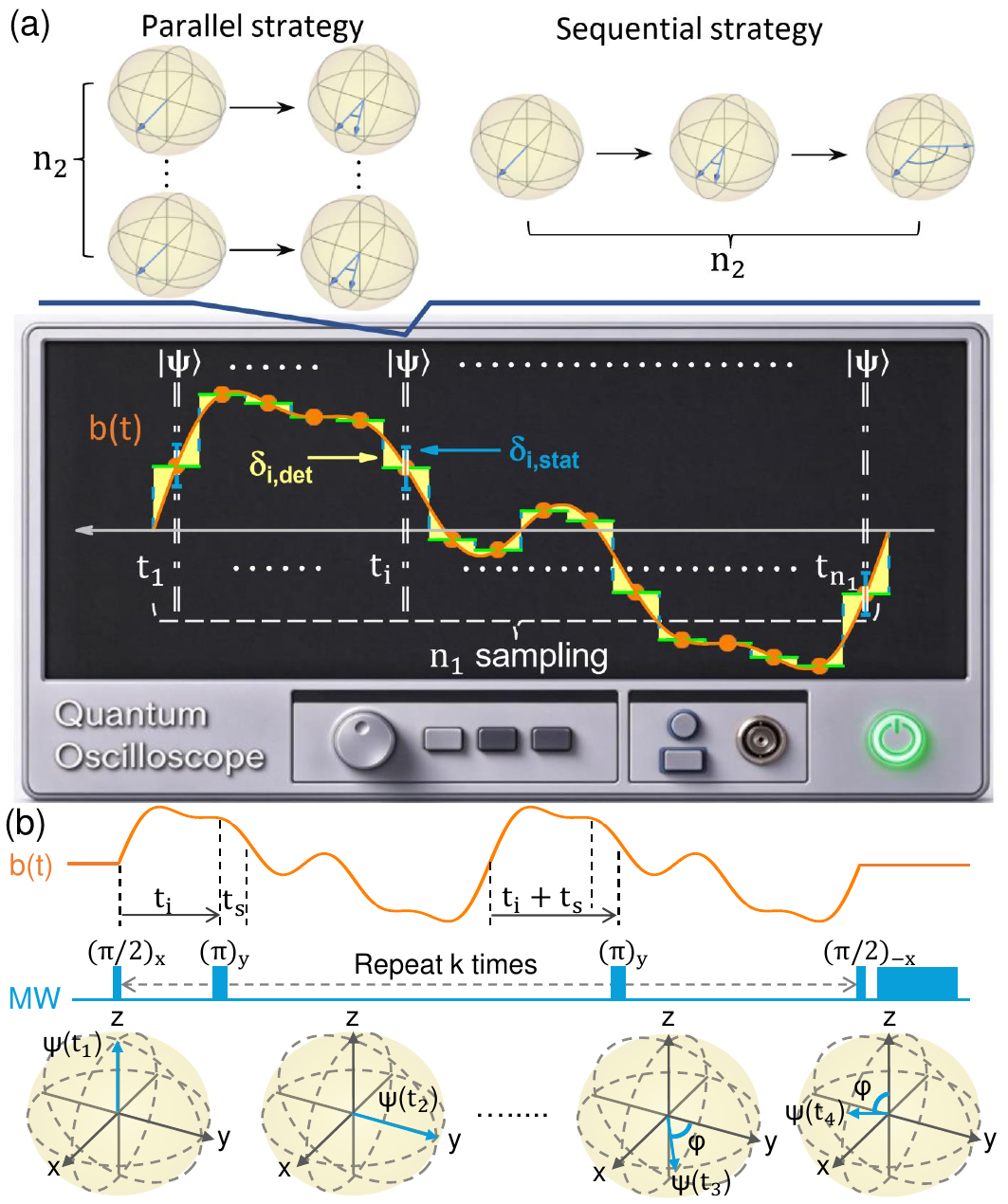}
\caption{Concept and mechanism of the Quantum Difference Amplification (QDA) method. (a) Comparison of the parallel entangled strategy and the sequential strategy. The detailed waveform reconstruction is conceptually illustrated via a ``Quantum Oscilloscope'' interface, highlighting the deterministic error \(\delta_{det}\) and statistical error \(\delta_{stat}\). (b) Microwave pulse sequence for the QDA protocol. By applying a series of \(\pi\) pulses, the unwanted phase accumulation outside the sensing window \([t_i, t_i + t_s]\) is canceled, while the target phase is coherently amplified. Bottom: Evolution of the probe state on the Bloch sphere during the sequence.}
\label{fig1}
\end{figure}

Quantum-enhanced sensing \cite{RevModPhys.90.035005,RevModPhys.90.035006,giovannetti2006quantum} exploits superposition and entanglement to yield higher statistical precision than purely classical approaches.  While the Heisenberg quantum limit (HQL) has been widely demonstrated in parameter estimation \cite{zhou2025entanglement,rovny2025multi}, extending these advantages to continuous time-dependent fields-categorized as waveform sensing \cite{tsang2011fundamental,herb2025quantum,duan2025concurrent,gianani2021kramers} is highly desirable to construct a quantum oscilloscope. Unlike finite-dimensional single-parameter estimation, waveform sensing is an infinite-dimensional function estimation problem, fundamentally distinguishing it from conventional quantum metrology. The first experimental demonstration beyond the SQL was reported by Yonezawa et al. \cite{yonezawa2012quantum}, who used squeezed light to track a stochastic optical phase. Subsequent theoretical work \cite{berry2015quantum,kura2020standard} further established that the overall waveform measurement error (\(\delta\)) is decomposed into two parts ${\delta ^2} = \delta _{det}^2 + \delta _{stat}^2,$ where ${\delta _{stat}}$ is the statistical error, and ${\delta _{det }}$ is the deterministic error as conceptually illustrated in Fig.~\ref{fig1}(a).

Specifically, for a waveform with H\"{o}lder continuity exponent \(q\), the deterministic error, arising from the finite sampling and reconstruction, scales as \(\delta_{det} = O(n_1^{-q})\) with respect to the number of sampling points \(n_1\). Meanwhile, the statistical error is dictated by the chosen measurement strategy: standard classical (or uncorrelated) measurement yields \(\delta_{stat} = O(n_2^{-1/2})\), while a quantum-enhanced method can suppress it to \(\delta_{stat} = O(n_2^{-1})\), where \(n_2\) is the number of resources allocated per sampling point. By optimizing the allocation of total quantum resources \(N = n_1 n_2\), the overall reconstruction error is fundamentally bounded by the SQL of \(\delta = O(N^{-q/(2q+1)})\) and the HQL of \(\delta = O(N^{-q/(q+1)})\). Therefore, the defining characteristic of a quantum oscilloscope is to achieve quantum-enhanced, time-domain waveform reconstruction, fundamentally surpassing the classical SQL in this infinite-dimensional function estimation.

In the canonical form of quantum-enhanced waveform sensing, achieving HQL scaling usually requires large-scale exotic entangled probes, which remains a non-trivial challenge with currently available quantum technology. As conceptually illustrated in Fig.~\ref{fig1}(a), unlike the parallel strategy that demands massive entangled resources, we employ a sequential sensing strategy where a single probe interacts with the time-dependent field repeatedly. Here, we implement this strategy by proposing a robust quantum difference amplification (QDA) method, which utilizes quantum differential detection to suppress the statistical error and achieve HQL scaling. As the quantum probe evolves continuously in the time-dependent field, unwanted phase accumulation is canceled via the quantum difference method instead of classical post-processing of the measurement data. In this way, the signals are amplified and error propagation is suppressed simultaneously by fully exploiting the quantum correlation resources. Using this QDA framework, we report the first experimental realization of a solid-state quantum oscilloscope. Specifically, within the zero-order hold (ZOH) reconstruction framework for waveforms with \(q=1\), our overall measurement error scales as \(O(N^{-1/2})\), unequivocally surpassing the classical \(O(N^{-1/3})\) limit and marking a genuine quantum advantage in continuous waveform reconstruction.

The detailed implementation of the QDA sequence is shown in Fig.~\ref{fig1}(b). By applying a flip \(P=(\pi)_y\) gate, the relative phase among probe states is exchanged. For \(t_s/T \ll 1\), the accumulated relative phase is expressed as \(\varphi (t_i) = \phi_1 - \phi_2 + \phi_3 \approx 2\gamma_e b(t_i)t_s\), assuming that the period \(T\) is precisely known. Crucially, this differential mechanism acts as a common-mode noise rejection process at the quantum level. By dynamically flipping the spin state with \(\pi\) pulses, the quantum probe effectively reverses the sign of the phase accumulation caused by background fluctuations. Consequently, any low-frequency environmental noise or control field drifts outside the narrow sensing window \([t_i, t_i + t_s]\) is symmetrically canceled out. This dynamic decoupling effect intrinsically protects the sensing process, ensuring that the extracted phase strictly originates from the target waveform within the sampling interval, thereby yielding a substantially enhanced signal-to-noise ratio (SNR) without requiring complex classical post-processing.

\begin{figure*}[tbp]
\centering
\includegraphics[width=0.75\textwidth]{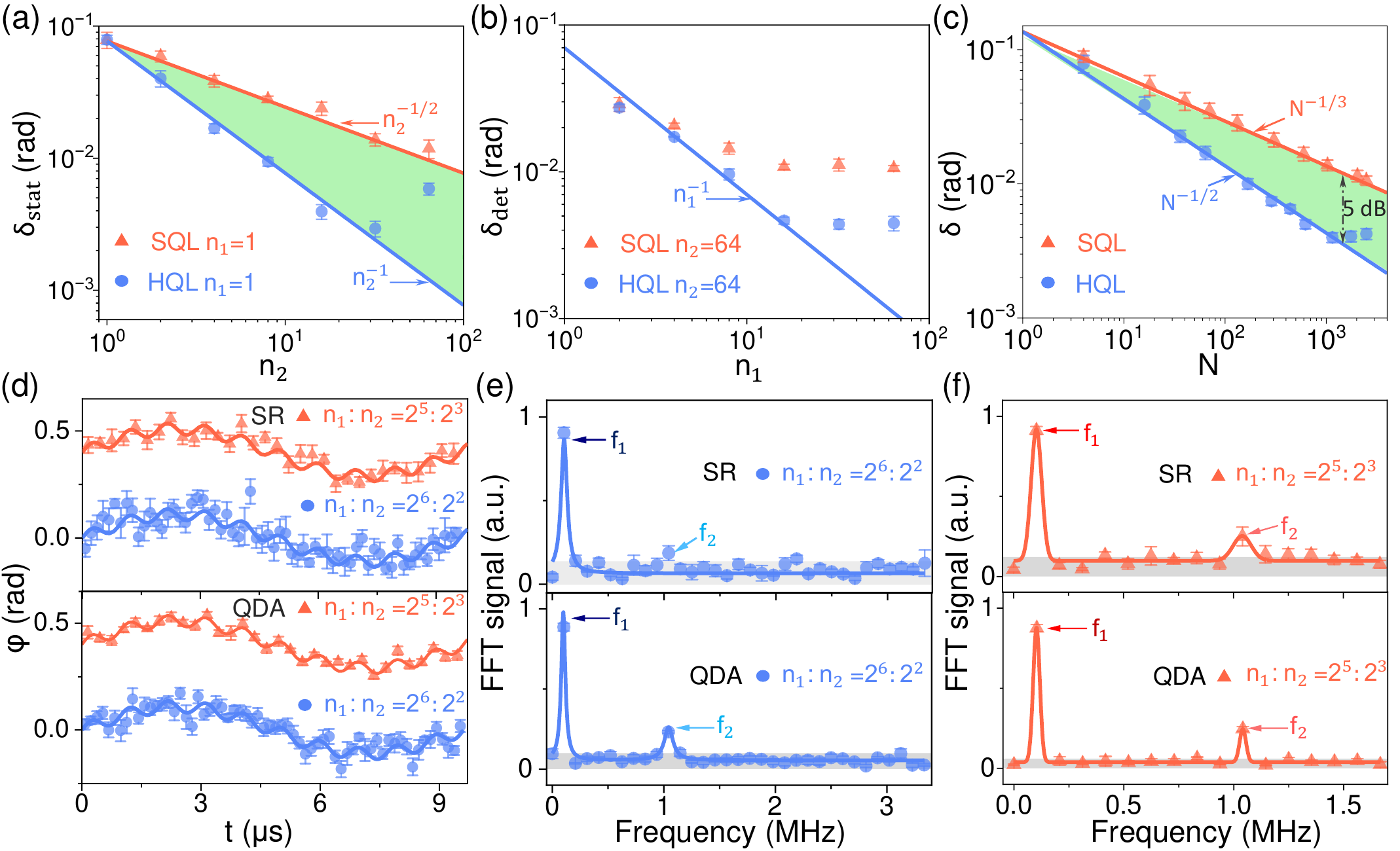}
\caption{Experimental verification of the QDA protocol. (a) Statistical error \(\delta_{stat}\) versus sequential resource \(n_2\), showing the transition from SQL (\(O(n_2^{-1/2})\)) to HQL (\(O(n_2^{-1})\)). (b) Deterministic error \(\delta_{det}\) versus sampling points \(n_1\). (c) Overall measurement error \(\delta\) versus total resource \(N\), demonstrating a 30-fold reduction in resource consumption to achieve a 5 dB error below the SQL. (d) Time-domain reconstruction of a complex waveform (${b(t) = b[sin(2}\pi {f_1}t) + {\text{sin(2}}\pi {f_2}t)/4]$) with equal total resources. (e)-(f) Fourier spectra under different resource allocations. The QDA method distinctly extracts the weak high-frequency signal \(f_2\), which is buried in shot noise for the SR method.}
\label{fig2}
\end{figure*}

We can control the repetition number \(k\) of the sequence to yield a sequential evolution. Hence, the accumulation relative phase is linearly increased as \(\varphi _{n_2} = k\varphi (t_i)\) and the unwanted phase accumulation is canceled out perfectly. The number of resources used for the QDA method is \(n_2=2k\) for each sampling. Benefiting from its optical spin polarization, efficient readout, and long room-temperature coherence time, we employ the nitrogen-vacancy (NV) center in diamond \cite{dong2024time,dong2023breaking} as a quantum probe. Furthermore, by incorporating dynamical decoupling techniques \cite{wang2023magnetic} during the free evolution periods, the effective pure dephasing time of the NV center is extended to \(T_2 \approx 640 \, \mu\text{s}\), ensuring the stability of the quantum probe throughout the extended sequential interactions.

For the same measurement precision, the QDA method provides a 30-fold reduction in quantum resource consumption compared with the conventional method. We test the basic scaling laws in the measurement of a sinusoidal waveform, where the standard small-interval Ramsey (SR) method is employed to avoid numerical processing for SQL. The experimental results are shown in Fig.~\ref{fig2}(a). The sensitivity of the unknown accumulation phase is improved by \(\sqrt{n_2}\)-fold compared with uncorrelated quantum resource, yielding the Heisenberg scaling. Results also show that when \(n_2 \ge 64\), \(\delta_{stat}\) gradually deviates from the theoretical HQL scaling. This deviation is caused by the transverse decoherence of the quantum probe, where the total coherent sensing time \(n_2T\) approaches the measured transverse coherence time \(T_2 \approx 64T\). Decoherence-induced statistical errors in the QDA method account for the observed departure from ideal HQL scaling.

In contrast, for the SR method, the measured \(\delta_{det}\) shows no clear scaling with \(n_1\) (Fig.~\ref{fig2}(b)), which arises because the deterministic error is quickly overwhelmed by the statistical shot noise floor. Under the SR protocol, the fixed resource allocation per sampling point anchors the statistical error at a constant floor, masking the theoretical \(O(n_1^{-1})\) scaling of the deterministic error. Furthermore, the finite repetition of experiments amplifies the shot noise fluctuations, making it challenging to resolve this scaling. However, the QDA scheme significantly suppresses the shot noise through coherent amplification, thereby unveiling the clear \(O(n_1^{-1})\) deterministic scaling. It is also worth noting that the signal contrast decays exponentially as \(A(n_2) = C \exp[-(n_2 T / T_2)^2]\), which accounts for the observed departure from the ideal HQL scaling at larger \(n_2\).

By making use of the inequality of arithmetic and geometric means, the trade-off between \(\delta_{stat}\) and \(\delta_{det}\) could be optimized numerically. Fig.~\ref{fig2}(c) shows the overall waveform measurement errors. Within the ZOH reconstruction framework, only \(\sim 1 \times 10^3\) quantum resources are required for the QDA method to achieve a waveform measurement error 5 dB below the SQL bound. In contrast, more than \(3 \times 10^4\) quantum resources are required for the conventional SR method, corresponding to a 30-fold reduction in resource consumption. As the ZOH scheme is currently the only reconstruction framework with a closed-form analytical expression, our observed \(q=1\) deterministic scaling provides a conservative error bound and a universal benchmark for any waveform with a bounded first derivative. Ultimately, these results experimentally demonstrate the unique measurement advantage of the quantum oscilloscope, marking the first experimental confirmation of this capability.

We further complete our study by demonstrating the measurement of a complex waveform, composed of different frequency components (\(f_1 = 0.104\) MHz and \(f_2 = 1.042\) MHz). The measured waveform with the SR and QDA method are shown in Fig.~\ref{fig2}(d). With the same number of resources, the statistical error is reduced significantly and the measurement results converge to the ideal value much faster with the QDA method. The fast Fourier transform (FFT) of measurement results is shown in Fig.~\ref{fig2}(e) and \ref{fig2}(f). For the same quantum resource, when the sampling rate is lower than twice the highest frequency of the signal, the QDA method enables high-SNR, high-resolution spectral measurement, even for weak high-frequency components that are unresolvable with the conventional SR method. This demonstrates that our method realizes the core waveform reconstruction function of a quantum oscilloscope within the ZOH framework. This capability is particularly crucial for nanoscale nuclear magnetic resonance (NMR) imaging, where signals from single nuclear spins are often extremely weak and easily buried in background noise. The QDA method enables the extraction of these weak, high-frequency spectral shifts, which can be used to precisely infer atomic positions and molecular connectivity. Furthermore, spectral addressability at this level is vital for operating large-scale quantum registers in solid-state quantum simulators.

In summary, we have experimentally demonstrated a room-temperature solid-state quantum oscilloscope utilizing the QDA method. The current results demonstrate the potential of our quantum measurement technique, which is readily transferable to other quantum probe platforms. Within the ZOH reconstruction framework, we experimentally demonstrate the SQL of \(O(N^{-1/3})\) and the HQL of \(O(N^{-1/2})\) with respect to the sequential quantum sensing resource. Our work is the experimental realization of a room-temperature solid-state quantum oscilloscope, extending quantum metrology from finite-dimensional parameter estimation to infinite-dimensional waveform sensing, and paving the way for practical quantum-enhanced time-domain measurement devices. While the current experiment focuses on deterministic and repeatable waveforms, our QDA framework can be readily extended to stochastic signals by integrating adaptive sampling strategies. Importantly, the fundamental resource equivalence between sequential coherence and multipartite entanglement implies that our method could be generalized to parallel entangled probe schemes. This allows for single-shot waveform reconstruction and asymptotic saturation of the quantum Cram\'{e}r-Rao bound, thereby expanding the application boundaries of quantum-enhanced metrology without changing the fundamental scaling laws demonstrated here.

\section*{Conflict of interest}
The authors declare they have no competing interests.

\section*{Acknowledgments}
This work was supported by  the National Natural Science Foundation of China (Grant Nos. 62225506, and 62305324). The sample preparation was partially conducted at the USTC Center for Micro and Nanoscale Research and Fabrication.

\section*{Author contributions}
Fang-Wen Sun and Yang Dong  conceived the idea of quantum oscilloscope, Yang Dong designed the QDA protocol, performed the experiments, analyzed the data, and wrote the manuscript. Wang Jiang fabricated diamond samples. Fang-Wen Sun supervised the project. All authors analyzed the data, discussed the results, and commented on the manuscript.

\section*{References}

\end{document}


\title{Supplementary Material for A Diamond Quantum Oscilloscope}
\author{Yang Dong}
\affiliation{{Laboratory of Quantum Information, University of Science and Technology of China, Hefei, 230026, P.R. China}}
\affiliation{{CAS Center For Excellence in Quantum Information and Quantum Physics, University of Science and Technology of China, Hefei, 230026, P.R. China}}
\affiliation{{Anhui Province Key Laboratory of Quantum Network, University of Science and Technology of China, Hefei, 230026, P.R. China}}
\affiliation{{School of Physics, Harbin Institute of Technology, Harbin 150006, P.R. China}}

\author{Yu Zheng}
\author{Wang Jiang}
\author{Zhi-Wei Liu}
\author{Shao-Chun Zhang}
\affiliation{{Laboratory of Quantum Information, University of Science and Technology of China, Hefei, 230026, P.R. China}}
\affiliation{{CAS Center For Excellence in Quantum Information and Quantum Physics, University of Science and Technology of China, Hefei, 230026, P.R. China}}
\affiliation{{Anhui Province Key Laboratory of Quantum Network, University of Science and Technology of China, Hefei, 230026, P.R. China}}
\author{Xiang-Dong Chen}
\author{Guang-Can Guo}
\affiliation{{Laboratory of Quantum Information, University of Science and Technology of China, Hefei, 230026, P.R. China}}
\affiliation{{CAS Center For Excellence in Quantum Information and Quantum Physics, University of Science and Technology of China, Hefei, 230026, P.R. China}}
\affiliation{{Anhui Province Key Laboratory of Quantum Network, University of Science and Technology of China, Hefei, 230026, P.R. China}}
\affiliation{{Hefei National Laboratory, University of Science and Technology of China, Hefei 230088, P.R. China}}

\author{Fang-Wen Sun}
\email{fwsun@ustc.edu.cn}
\affiliation{{Laboratory of Quantum Information, University of Science and Technology of China, Hefei, 230026, P.R. China}}
\affiliation{{CAS Center For Excellence in Quantum Information and Quantum Physics, University of Science and Technology of China, Hefei, 230026, P.R. China}}
\affiliation{{Anhui Province Key Laboratory of Quantum Network, University of Science and Technology of China, Hefei, 230026, P.R. China}}
\affiliation{{Hefei National Laboratory, University of Science and Technology of China, Hefei 230088, P.R. China}}

\maketitle

In the supplementary material, we provide the theoretical derivations and experimental details for the diamond quantum oscilloscope. We first derive the decomposition of the waveform measurement error into statistical and deterministic components and determine the optimal resource allocation to achieve Heisenberg scaling. We then detail the room-temperature NV-center platform, the pulse sequences for both standard Ramsey and QDA protocols, and a quantitative analysis of the period mismatch tolerance. Finally, we present additional experimental data verifying the sequential phase accumulation and demonstrating that the QDA method resolves weak high-frequency components under frequency aliasing, which are inaccessible to conventional sampling.

\begin{flushleft}
\bf{S1. Waveform sensing theory}
\end{flushleft}

With the standard measurement scheme, we detect the unknown waveform based on \(N = n_1 n_2\) measurement results. In the present case, the overall results can be divided into \(n_1\) equal parts according to sampling time. For the sampling instant \(t_i\), we have a serial output \(\{\phi_{i,j}\}\) to get the unknown phase, where \(i = 1, 2, \ldots, n_1\) and \(j = 1, 2, \ldots, n_2\). Hence, the total measurement error of waveform has a simple form:
\begin{equation}
\begin{aligned}
\delta^2 &= E\left[\int_0^T \frac{dt}{T} [\bar{\phi}(t) - \phi(t)]^2 \right] \\
&= \frac{1}{n_2}\sum_{j=1}^{n_2}\sum_{i=1}^{n_1}\int_{t_i - \frac{T}{2n_1}}^{t_i + \frac{T}{2n_1}}\frac{dt}{T} [\phi_{i,j} - \phi(t)]^2 \\
&= \sum_{i=1}^{n_1}\left[ \frac{1}{n_2}\sum_{j=1}^{n_2}\int_{t_i - \frac{T}{2n_1}}^{t_i + \frac{T}{2n_1}}\frac{dt}{T} [\phi_{i,j} - \bar{\phi}_i + \bar{\phi}_i - \phi(t)]^2 \right],
\end{aligned}
\end{equation}
where we define \(\bar{\phi}_i = \frac{1}{n_2}\sum_{j=1}^{n_2}\phi_{i,j}\) based on the least square method. In this way, the cross-term is zero:
\begin{equation}
\sum_{i=1}^{n_1}\left[ \frac{1}{n_2}\sum_{j=1}^{n_2}\int_{t_i - \frac{T}{2n_1}}^{t_i + \frac{T}{2n_1}}\frac{dt}{T} 2(\phi_{i,j} - \bar{\phi}_i)(\bar{\phi}_i - \phi(t)) \right] = 0.
\end{equation}

And we have
\begin{equation}
\begin{aligned}
{\delta ^2} =& \sum\nolimits_{i = 1}^{{n_1}} {\left[ {\frac{1}{{{n_2}}}\sum\nolimits_{j = 1}^{{n_2}} {\int_{{t_i} - \frac{T}{{2{n_1}}}}^{{t_i} + \frac{T}{{2{n_1}}}} {\frac{{dt}}{T}} } {{\left( {{\phi _{i,j}} - {{\bar \phi }_i}} \right)}^2} + \frac{1}{{{n_2}}}\sum\nolimits_{j = 1}^{{n_2}} {\int_{{t_i} - \frac{T}{{2{n_1}}}}^{{t_i} + \frac{T}{{2{n_1}}}} {\frac{{dt}}{T}} } {{\left( {{{\bar \phi }_i} - \phi \left( t \right)} \right)}^2}} \right]} \\
=& \sum\nolimits_{i = 1}^{{n_1}} {\left[ {\int_{{t_i} - \frac{T}{{2{n_1}}}}^{{t_i} + \frac{T}{{2{n_1}}}} {\frac{{dt}}{T}} \delta _{i,stat}^2 + \int_{{t_i} - \frac{T}{{2{n_1}}}}^{{t_i} + \frac{T}{{2{n_1}}}} {\frac{{dt}}{T}} \delta _{i,det}^2} \right]} \\
=& \frac{1}{{{n_1}}}\sum\nolimits_{i = 1}^{{n_1}} {\delta _{i,stat}^2 + \int_0^T {\frac{{dt}}{T}} \delta _{i,det}^2} \\
=& \delta _{stat}^2{\rm{ + }}\delta _{det}^2 \text{,}
\end{aligned}
\end{equation}
where \(\delta_{i,stat}^2 = \frac{1}{n_2}\sum_{j=1}^{n_2}(\phi_{i,j} - \bar{\phi}_i)^2\) and \(\delta_{i,det}^2 = [\bar{\phi}_i \theta(t - t_i + \frac{T}{2n_1}) - \phi(t)]^2\). Hence, the overall waveform measurement error can be decomposed into two parts: the statistical error and the deterministic error. As \(n_2\) increases large enough, the average measurement result \(\bar{\phi}_i\) would converge to the ideal value due to the central limit theorem. And for a finite number of experiment repeats, there will be some of the discrepancy between theory and experiment. So once the average measurement results \(\bar{\phi}_i\) converges to the ideal values of arbitrary waveform, the deterministic error scales as:
\begin{equation}
\begin{aligned}
\delta_{det}^2 &= \sum_{i=1}^{n_1}\int_{t_i - \frac{T}{2n_1}}^{t_i + \frac{T}{2n_1}} \left[ \bar{\phi}_i \theta(t - t_i + \frac{T}{2n_1}) - \phi(t) \right]^2 \\
&= \sum_{i=1}^{n_1}\int_{t_i - \frac{T}{2n_1}}^{t_i + \frac{T}{2n_1}}\frac{dt}{T} O(t^{2q}) \\
&= O(n_1^{-2q}).
\end{aligned}
\end{equation}
For the ZOH reconstruction method used in this work, the deterministic error scaling \(\delta_{det} = O(n_1^{-q})\) follows from the H\"{o}lder continuity assumption \(|\phi(t + \epsilon) - \phi(t)| = O(\epsilon^q)\) for \(0 < q \le 1\). For waveforms that are smoother than the \(q=1\) class (e.g., analytic functions such as sinusoids), the ZOH reconstruction error is conservatively bounded by the \(q=1\) scaling \(\delta_{det} = O(n_1^{-1})\). This is the appropriate error bound for the ZOH method, which is the only reconstruction scheme in the theoretical framework with a closed-form analytical expression. For \(q>1\), the Kura-Ueda framework proves the existence of optimal reconstruction functions but does not provide explicit closed-form expressions. Our experimental verification of the \(q=1\) scaling thus represents the first experimentally realizable demonstration of quantum-enhanced waveform reconstruction within this framework.

For a given number of resources \(N = n_1 n_2\), the optimal measurement is determined by the trade--off between these errors. By employing the inequality of arithmetic and geometric means, the minimum measurement error of waveform is:
\begin{equation}
\begin{aligned}
\delta^2 &=\delta_{det}^2+ \delta_{stat}^2 \\
&= O(M_1n_1^{-2}) + O(n_2^{-1})
\end{aligned}
\end{equation}
in SR method. If we neglect the constant factor, we will get the optimal scaling relationship \(\delta = O(N^{-1/3})\) in the SQL scheme. With similar procedure, we have
\begin{equation}
\begin{aligned}
\delta^2 &=\delta_{det}^2+ \delta_{stat}^2 \\
&= O(M_2n_1^{-2}) + O(n_2^{-2})
\end{aligned}
\end{equation}
and the minimum measurement error scale as \(\delta = O(N^{-1/2})\) for the QDA method. The optimal quantum resource allocation of arbitrary waveform estimation would meet this condition \(n_1^{-2} \sim n_2^{-1}\) for SQL and \(n_1^{-2} \sim n_2^{-2}\) for HQL with large \(N\). And this relationship has a simple form: \(\log_2 n_1 = \frac{1}{2}\log_2 n_2 + c_1\) and \(\log_2 n_1 = \log_2 n_2 + c_2\) for SQL and HQL, respectively.

\begin{flushleft}
\bf{S2. NV center in diamond and measurement setup}
\end{flushleft}

The NV center in diamond, where a substitutional nitrogen atom is next to a vacancy, forms a spin triplet system in its ground state. The NV's spin sub-levels of ground state are \(|0\rangle\) and \(|\pm 1\rangle\). The zero-field splitting is \(D = 2.87\) GHz. In the experiment, we apply an external magnetic field along the NV symmetry axis to polarize intrinsic \(^{14}\)N nuclear spin to enhance the signal contrast up to \(C = 0.3\). Electron spin states \(|0\rangle\) and \(|1\rangle\) of NV center are encoded as the probe qubit. We use the home-built confocal microscope to image, initialize and read out NV center in a single--crystal synthetic diamond sample (Element Six). The NV centers are generated by 40 keV $^{14}{N^ + }$ ion implantation followed by annealing in high vacuum at 800 $^0$C. The boiling tri-acid solution (a balanced mix of nitric, sulfuric, and perchloric acids) is used to clean the diamond sample and is eventually diluted by deionized water. The estimated average depth of the NV is about 40 nm.
\begin{figure}[htbp]
\centering
\includegraphics[width=0.8\textwidth]{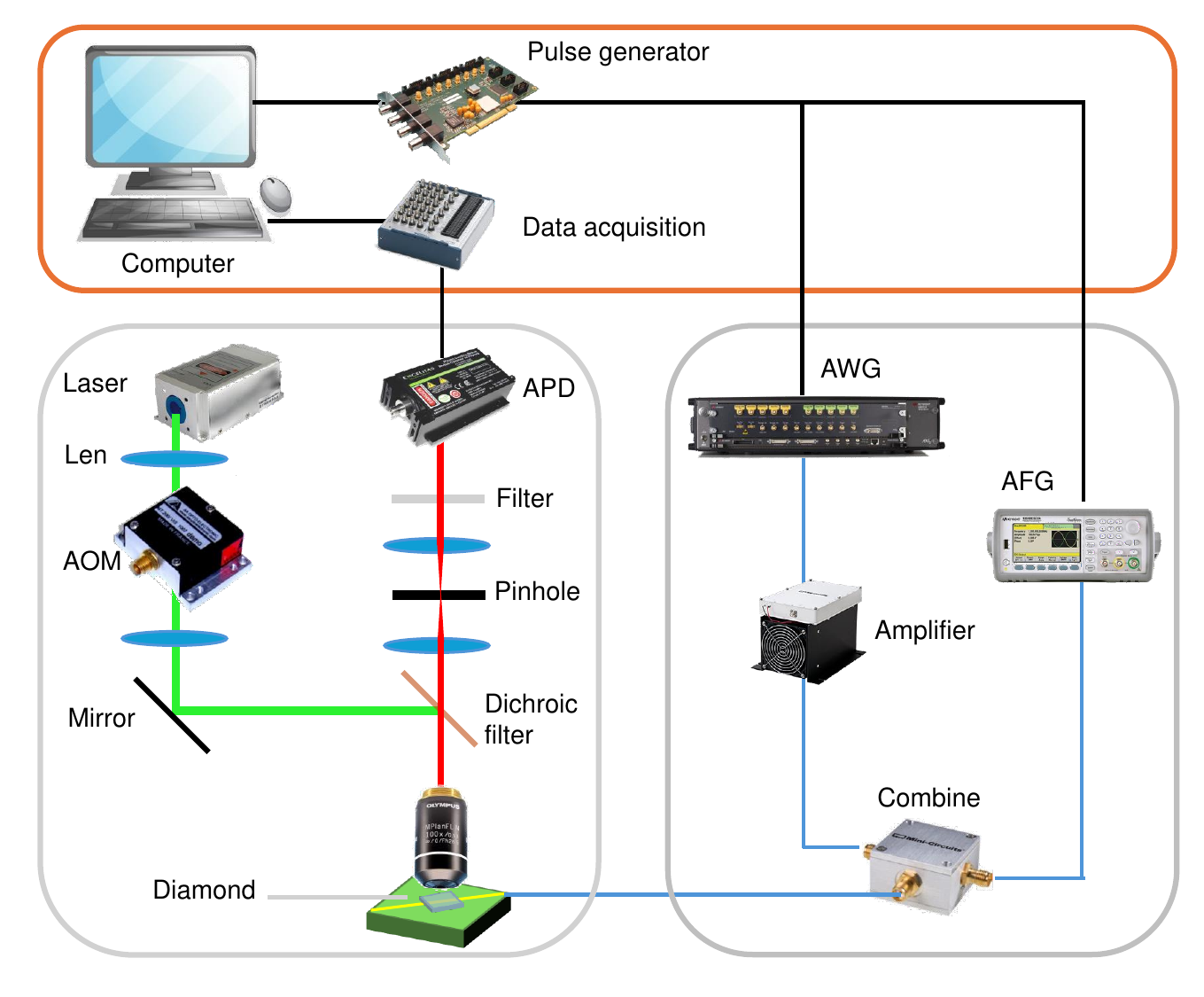}
\caption{The experimental platform. The platform consists of an optical system, a microwave (MW) system and a signal processing system, as well as a piezoelectric translation stage and a permanent magnet below the sample.}
\label{fig:s1}
\end{figure}

Fluorescence photons are collected into a fiber and detected by the single photon counting module (SPCM-AQRH-15-FC, Excelitas), with a counting rate of 70 kHz and a signal-to-noise ratio of 50:1. It is separated from the excitation light by a Semrock dichroic mirror, a long-pass filter with a 650 nm cutoff, and spatially filtered through a 15 \( \mu\text{m}\) pinhole. The MW pulses for controlling the NV center are generated by an arbitrary waveform generator (AWG, Keysight M8190A) and fed into the coplanar waveguide microstructure on the quartz glass. The driving MW is amplified by an amplifier (ZHL-25W-63+, Mini-Circuits). A SpinCore programmable pulse generator (PulseBlaster ESR-PRO 500, SpinCore) controls the MW switch and acousto-optic modulator (AOM) as shown in Fig. \ref{fig:s1}. The position of the NV center is adjusted using a piezo stage (P-611.3, Physik Instrument).

In the experiment, the observed fluorescence photon number $I$, which is related to the spin population distributions between the $m_s=0$ and $m_s=+1$ states of the NV center, can be converted to the probability of quantum coherent operation by linear transformation. Specifically, we firstly initialize the system into the $m_s=0$ state by a 532 nm laser pulse and then change it into the $m_s=+1$ state by MW operation and measured their photon counts (denoted by $I_{max }, I_{min}$). In order to beat the fluctuation of photon counting, we repeat the experimental cycle more than $10^5$ times. So the relative population of the $m_s=0$ state for an unknown state can be expressed as $P_0=\frac{I-I_{min }}{I_{max }-I_{min }}$. Therefore, the quantum state of the NV center electron spin can be determined from the fluorescence intensity of NV center.

Under resonance condition, the Rabi frequency is set to be \(\Omega = 20\) MHz in experiment. The undesired couplings between the NV center and the surrounding \(^{13}\)C nuclear spins will lead to the effective dephasing effect. By making use of the dynamical decoupling, the truly pure dephasing time \(T_2\) is extended to 640 \( \mu\text{s}\). The decoherence behavior of NV center is non-trivial and recent studies have shown that the sources of decoherence noise in NV centers is classified by frequency \cite{han2025solid}. Below 1 kHz, the electronic spin bath is the main contributor to the noise spectrum, showing that the interactions between the central spin and its specific surrounding electronic spin bath play a crucial role in spin decoherence. Above 1 MHz, the noise spectrum is dominated by the intrinsic spin relaxation, suggesting that the relaxation of spins to their equilibrium state fundamentally determines the coherence physical limit of a spin qubit within a quantum system. Between 1 kHz and 1 MHz, the spectrum is primarily influenced by a noise source with a high cutoff frequency around 1 MHz, which is treated as an emerging noise mechanism. And the interval of the MW pulses is about 10 $\mu s$ (i.e., 100 kHz), and thus the emergent noise mechanism dominates the effect in our work.

\begin{flushleft}
\bf{S3. The small-interval Ramsey method}
\end{flushleft}

\begin{figure}[bp]
\centering
\includegraphics[width=0.5\textwidth]{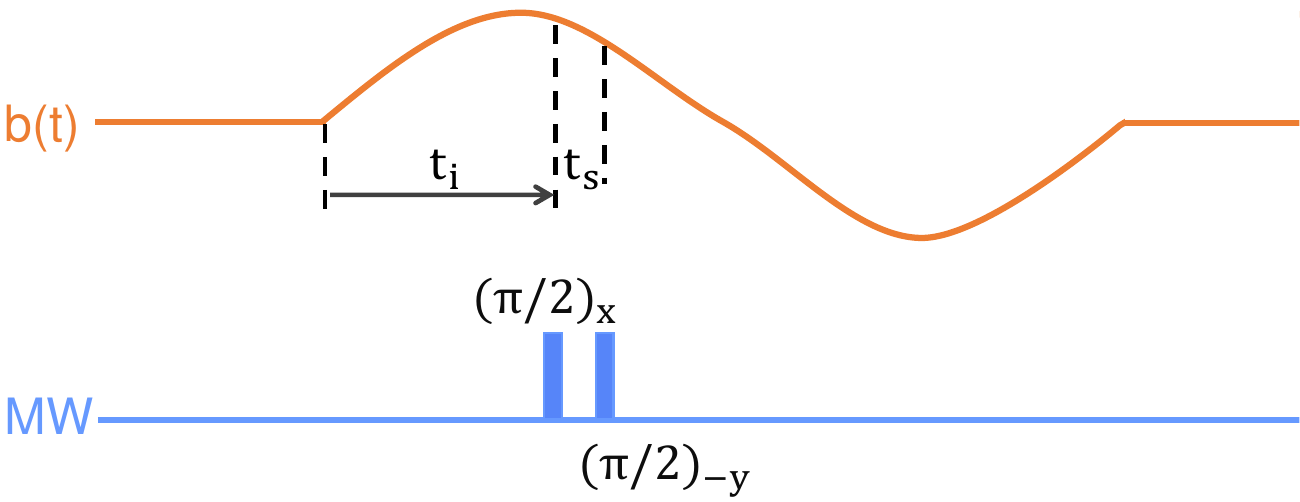}
\caption{The SR method for SQL waveform measurement. Pulse sequence for the SQL measurement of waveform sampling. We repeat this sensor sequence \(2k\) times to match the total duration of a QDA sequence with \(k\) repetitions as shown in Fig. 1(b) of the main text.}
\label{fig:s2}
\end{figure}

In a small-interval Ramsey (SR) experiment, a superposition state, prepared by a first \((\pi/2)_x\) gate at \(t_i\), evolved during a measurement time \(t_s\) and acquired a phase factor \(\phi(t_i) = \int_{t_i}^{t_i+t_s}\gamma_e b(t)dt \approx \gamma_e b(t_i)t_s\). To extract the phase, \(\phi(t_i)\) is typically converted into a population difference \(p(t_i)\) by a second \((\pi/2)_x\) gate, \(p(t_i) = \frac{1}{2}[1 + \sin\phi(t_i)] \approx \frac{1}{2}[1 + \phi(t_i)]\) with \(\phi(t_i) \ll 1\), followed by a projection measurement of the probe and signal averaging. By employing this method, we realize the sampling of the waveform in small intervals and built up the waveform by stepping \(t_i\) as shown in Fig. \ref{fig:s2}.

\begin{flushleft}
\bf{S4. Detailed implementation of the QDA sequence}
\end{flushleft}

\begin{figure}[bp]
\centering
\includegraphics[width=0.95\textwidth]{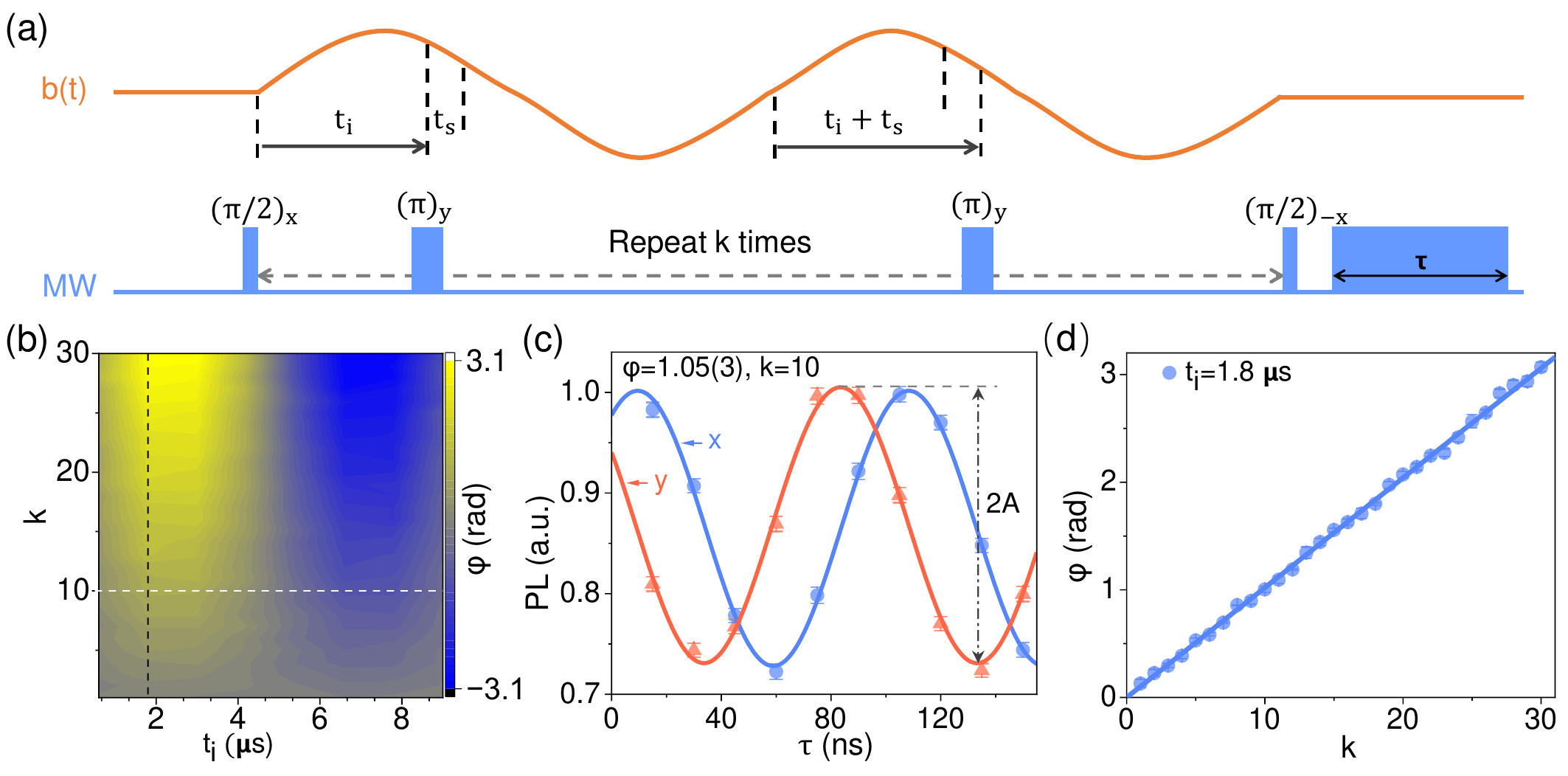}
\caption{Verification of the QDA sequence. (a) The pulse sequence of the QDA method. (b) The detection of a sinusoidal waveform \(b(t) = b\sin(2\pi ft)\) with a period of \(T = 9.6 \) \(\mu\text{s}\). The readout of accumulation phase from a sinusoidal test signal recorded with different values of \(t_i\) and \(k\). The integration time is \(t_s = 50\) ns. (c) The unknown phase $\varphi $ is mapped into the initial phase of Rabi oscillation and read by optical method (probing along \(x\) and \(y\) axes) with \(k=10\) and \(t_i = 1.8 \) \(\mu\)s, corresponding to the intersection of white and black dashed lines in (b). (d) The accumulation phase $\varphi $ as a function of repeat number \(k\) with a fixed \(t_i = 1.8 \) \(\mu\)s. The theoretical solid line shows a linear scaling.}
\label{fig:s3}
\end{figure}

Our method follows a general sequence of quantum probe initialization, interaction with the waveform, projection readout. The detailed implementation is shown in  Fig. \ref{fig:s3}(a):

(1) The quantum probe is initialized into a known ground state, for example, $\left| 0 \right\rangle $.

(2) The quantum probe is transformed into the state of $\left| \psi  \right\rangle {\text{ = }}\frac{{\left| 0 \right\rangle  + \left| 1 \right\rangle }}{{\sqrt 2 }}$ with ${\left( {\pi /2} \right)_x}$ gate.

(3) The superposition state evolves under the Hamiltonian \(H = \gamma_e b(t)\hat{S}_z\). Such evolution is dictated by the unitary transformation \(U_1 = e^{-i\int_0^t Hdt}\) and the probe state after the evolution is \(|\psi_1\rangle = \frac{|0\rangle + e^{-i\phi_1}|1\rangle}{\sqrt{2}}\). The relative phase is ${\phi _1}({t_i}) = \int_0^{{t_i}} {{\gamma _e}b(t)dt} $.

(4) By applying a flip \(P = (\pi)_y\) gate, the relative phase among probe state is exchanged \(|\Psi_1\rangle = \frac{e^{-i\phi_1}|0\rangle + |1\rangle}{\sqrt{2}}\).

(5) The superposition state evolves under the Hamiltonian \(H\) for a time \(T+t_s\). Such evolution is dictated by the unitary transformation \(U_2\) and the probe state after the evolution is \(|\psi_2\rangle = \frac{e^{-i(\phi_1-\phi_2)}|0\rangle + |1\rangle}{\sqrt{2}}\).

(6) By applying a flip \(P = (\pi)_y\) gate, the relative phase among probe state is exchanged \(|\Psi_2\rangle = \frac{|0\rangle + e^{-i(\phi_1-\phi_2)}|1\rangle}{\sqrt{2}}\).

(7) The superposition state evolves under the Hamiltonian \(H\) for a time \(T+t_s-t_i\). The probe state after the evolution is \(|\psi_3\rangle = \frac{|0\rangle + e^{-i(\phi_1-\phi_2+\phi_3)}|1\rangle}{\sqrt{2}}\). For \(t_s/T \ll 1\), the accumulated relative phase is expressed as: $\varphi  ({t_i}) = \int_0^{{t_i}} {{\gamma _e}b(t)dt}  - \int_{{t_i}}^{T + {t_i} + {t_s}} {{\gamma _e}b(t)dt}  + \int_{T + {t_i} + {t_s}}^{2T} {{\gamma _e}b(t)dt}  =  - 2\int_{{t_i}}^{{t_i} + {t_s}} {{\gamma _e}b(t)dt}  \approx 2{\gamma _e}b({t_i}){t_s}$.

(8) Using a second \((\pi/2)_{-x}\) or \((\pi/2)_-y\) gate, the relative phase $\varphi $ is mapped into \(x-z\) or \(y-z\) plane of Bloch ball.

(9) The accumulated relative phase \(\varphi\) is transferred to the initial phase of NV electron-spin Rabi oscillation, rather than being directly read out. We sweep the MW driving time, record the resulting PL oscillation curves, and perform sinusoidal fitting to extract this initial phase, which gives the value of the target sensing phase \(\varphi\). This Rabi-oscillation-based phase readout provides robust phase estimation in our experiment.

Steps 2-9 represent a single measurement cycle. We can control the repetition number \((k)\) of steps 3-7 to yield a sequential evolution. Hence, the accumulation relative phase is linearly increased as \(\varphi_{n_2} = k\varphi (t_i)\) and the unwanted phase accumulation for the rest interaction time is canceled out perfectly. The number of resources used for the QDA method is ${n_2} = 2k$ for each sampling. And the $\pi$ gate is composite pulses \cite{PhysRevB.103.104104}: $\pi_y=e^{-i \sigma_y \Omega \frac{\tau}{2}} e^{-i \sigma_y \frac{\Omega}{2} \tau} e^{-i \sigma_y \Omega \frac{\tau}{2}}$ and $\Omega \tau=\pi$. The detuning effect ($\Delta$) induced by the waveform $b(t)$ will be suppressed up to fourth order $O\left((\Delta / \Omega)^4\right)$. For room-temperature NV center quantum sensing, the maximum Rabi frequency of the target signal is restricted to below 100 MHz by the limited MW manipulation rate, while the minimum frequency is constrained to above 3 kHz by the NV center coherence time ($\sim$1 ms). The experimentally accessible signal bandwidth is therefore approximately 5 kHz-100 MHz. Hence, combined with the preliminary rough measurement results, our method is applicable as long as the frequency of the repeatable waveform signal falls within this range.

To ensure a fair comparison, we repeat the SR method \(2k\) times, matching the \(2T\) total duration of a QDA sequence with \(k\) repetitions, thereby reducing signal fluctuations and achieving the SQL in waveform estimation. For the QDA method, we initialize the NV spin probe, apply ${n_2} = 2k$ concatenated repetitive waveform signal and then perform a single projective measurement of the spin state using optical readout. We select ${n_1}$ distinct measurement time points $\{ {{t}_1}{\text{,}}{{t}_2}{\text{,}}{{t}_3}{\text{,}} \cdots \} $. To improve the signal-to-noise ratio (SNR) of the detection results, the above whole process was repeated $R = 1.2 \times {10^5}$ times. For SR method, we apply a single Ramsey gate (interaction with signal waveform one time), perform $2{n_2}$ projective measurement to estimate the phase. We select ${n_1}$ distinct measurement time points $\{ {{t}_1}{\text{,}}{{t}_2}{\text{,}}{{t}_3}{\text{,}} \cdots \} $. To improve the SNR of the detection results, the above whole process was also repeated $R$ times. Therefore, the amount of resources consumed by the two methods is the same.

\begin{flushleft}
\bf{S5. Quantitative analysis of residual phase error caused by imperfect $T$}
\end{flushleft}

We note that the cancellation of unwanted phase accumulation in the QDA protocol relies on the assumption that the waveform is repeatable and that its period $T$ is known or can be pre-determined with sufficient accuracy. For a periodic signal with a known period $T$, the protocol achieves perfect differential cancellation of the phase accumulated outside the sensing window ${[}{{t}_i}{\text{, }}{{t}_i}{ + }{{t}_s}{]}$. Now, we first quantitatively derive the residual phase error induced by period mismatch $\Delta T$ : $\varphi \left( {{t_i},\Delta T} \right) = \int_0^{{t_i}} {{\gamma _e}} b(t)dt - \int_{{t_i}}^{T + \Delta T + {t_i} + {t_s}} {{\gamma _e}} b(t)dt + \int_{T + \Delta T + {t_i} + {t_s}}^{2T + 2\Delta T} {{\gamma _e}} b(t)dt \approx 2{\gamma _e}b\left( {{t_i}} \right){t_s} + 2{\gamma _e}\left[ {b(t = 0) - b\left( {t = {t_i} + {t_s}} \right)} \right]\Delta T$. We define the relative phase deviation $\delta_{\varphi}$ as the fractional error between the distorted phase and ideal phase $(\Delta T=0)$:
${\delta _\varphi } = \left| {\frac{{\varphi \left( {{t_i},\Delta T = 0} \right) - \varphi \left( {{t_i},\Delta T} \right)}}{{\varphi \left( {{t_i},\Delta T = 0} \right)}}} \right| \approx \frac{{2T}}{{{t_s}}}{\varepsilon _T}$,
and $\varepsilon_T=\frac{\Delta T}{T} \approx \frac{\Delta f}{f}=\varepsilon_f$ when $\Delta T \ll T$. The claimed scaling relationship will be kept if $\delta_{\varphi} \ll \delta$. Here, the $\delta$ is total measurement error of waveform in our work. This yields a necessary tolerance condition for frequency/period estimation: $\varepsilon_{\mathrm{f}} \approx \varepsilon_{\mathrm{T}} \ll \frac{t_{\mathrm{s}}}{2 T} \delta$. For the core scaling and signal-to-noise ratio (SNR) benchmark experiments presented in the manuscript, the signal period $T$ is precisely predefined and calibrated as an input parameter.

For the scaling demonstrations and SNR benchmark experiments reported in this manuscript, the waveform frequencies are controlled and calibrated experimentally. Therefore, the characteristic period $T$ is known with accuracy much higher than the tolerance required by the QDA protocol. For practical measurements on unknown periodic signals, high-precision determination of $T$ can be realized by combining the SR sampling with quantum heterodyne (Qdyne) detection \cite{schmitt2017submillihertz,boss2017quantum}, as established theoretically and experimentally in prior landmark works. The main idea of the Qdyne detection scheme shifts the fundamental limit of frequency resolution in quantum spectroscopy from the qubit coherence time to the long-term stability of an external classical clock, and breaks the intrinsic resolution bound via a hybrid architecture combining short-timescale quantum sampling with long-duration classical spectral analysis. Hence, for the detection of period signal with quantum sensor, the period $T$ can be treated as a pre-known parameter or measured with high precision (which means $\varepsilon_{f} \approx \varepsilon_{T} \ll \frac{t_{s}}{2 T} \delta$ can be held for a periodic signal with unknown period in experiment). Such frequency characterization techniques provide a possible route for obtaining the required temporal scale before QDA sensing, although this preprocessing step is beyond the scope of the present demonstration. We focus on demonstrating that QDA method can achieve detection results with a high SNR within identical measurement durations. We further clarify that the present work focuses on deterministic and repeatable waveform reconstruction rather than completely unknown stochastic signals. Adaptive strategies, such as coarse spectral scanning or multi-scale sensing, may provide possible extensions toward more general sensing scenarios, which we now discuss as future directions rather than demonstrated capabilities.

\begin{flushleft}
\bf{S6. Verification of sequential phase accumulation}
\end{flushleft}

We investigate the experimental result by accumulating phase from \(n_2=2k\) consecutive waveform passages. Fig. \ref{fig:s3}(b) plots the readout of accumulation phase from a sinusoidal test signal recorded with different values of \(k\). With the increase of \(k\), a much stronger oscillation response is observed with a much higher dynamic range. Fig. \ref{fig:s3}(c) shows the readout process of an unknown accumulation phase. Furthermore, Fig. \ref{fig:s3}(d) shows the accumulation phase is strictly proportional to \(k\). Without correcting the decoherence of quantum probe, we observe the exact linear scaling relationship. Hence, the core of the QDA method coherently amplifies the accumulated phase via the sequential strategy and cancels unwanted evolution. When the waveform is rectangular function, $\pi $ gates in the sequence are not required and the QDA method will turn out to be the multi-pass protocol.

\begin{figure}[tbp]
\centering
\includegraphics[width=0.75\textwidth]{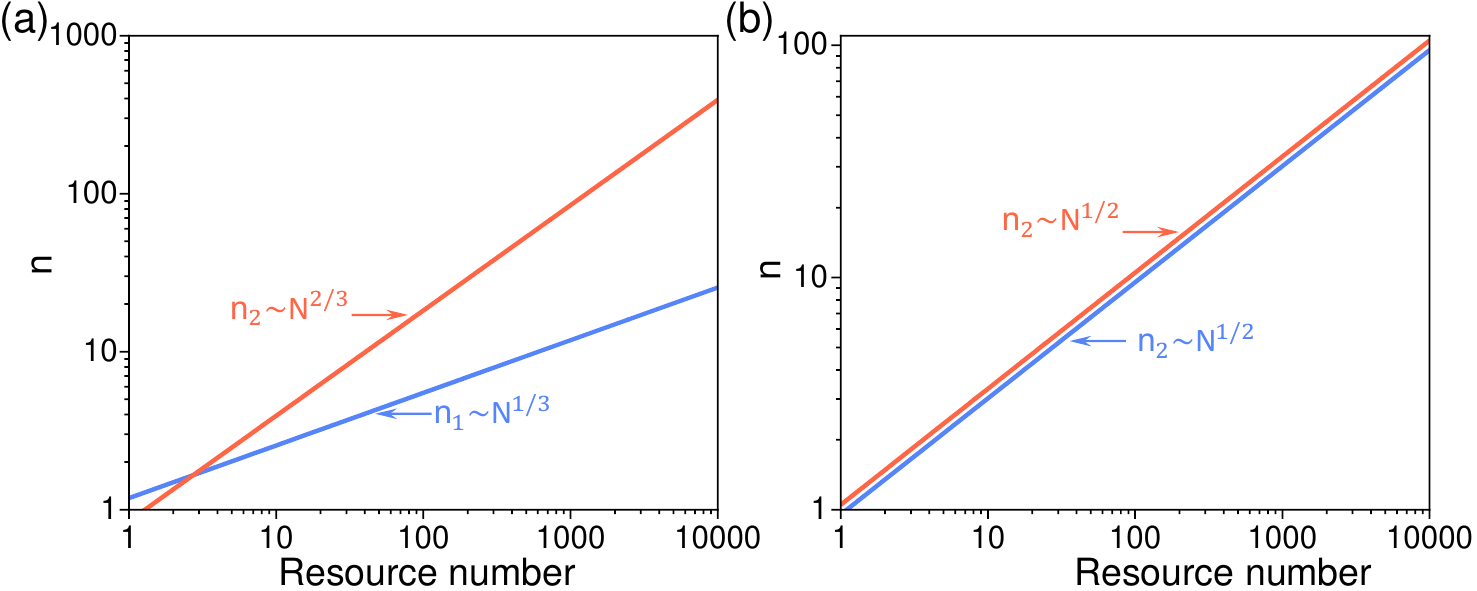}
\caption{Optimal quantum resource allocation for waveform reconstruction within ZOH reconstruction framework. (a)(b) The best allocation of quantum resource (\(n_1\) and \(n_2\)) for waveform sensing with SQL or HQL scheme within the zero--order hold (ZOH) reconstruction framework, respectively, as a function of total resource number \(N\).}
\label{fig:s4}
\end{figure}

\begin{flushleft}
\bf{S7. Numerical optimal quantum resource allocation}
\end{flushleft}

By fitting experimental results in Fig. 2(a) and Fig. 2(b) of the main text, the scaling laws show \(\delta_{stat} = 0.077/\sqrt{n_2}\) (\(\delta_{stat} = 0.077/n_2\)) for SQL (HQL) and \(\delta_{det} = 0.07/n_1\). Since the overall error is \(\delta^2 = \delta_{det}^2 + \delta_{stat}^2\), the best allocation of quantum resource for the waveform sensing can be calculated with numerical method and results are shown in Fig. \ref{fig:s4}. In the experiment, we adopt a numerical approximate rounding method to get the best quantum resource allocation approximately as shown in Tables \ref{tab:s1} and \ref{tab:s2}.

\begin{table}[htbp]
\centering
\caption{Optimal resource allocation for the small-interval Ramsey method within the ZOH reconstruction framework.}
\label{tab:s1}
\begin{tabular}{c|cccccccccc}
\hline
NO. & 1 & 2 & 3 & 4 & 5 & 6 & 7 & 8 & 9 & 10 \\
\hline
\(N\) & 4 & 18 & 40 & 70 & 132 & 304 & 600 & 1032 & 2040 & 2496 \\
\(n_1\) & 1 & 2 & 3 & 4 & 5 & 6 & 8 & 10 & 12 & 15 \\
\(n_2\) & 2 & 6 & 10 & 14 & 38 & 38 & 60 & 86 & 136 & 156 \\
\hline
\end{tabular}
\end{table}

\begin{table}[htbp]
\centering
\caption{Optimal resource allocation for the QDA method within the ZOH reconstruction framework.}
\label{tab:s2}
\begin{tabular}{c|ccccccccccc}
\hline
NO. & 1 & 2 & 3 & 4 & 5 & 6 & 7 & 8 & 9 & 10 & 11 \\
\hline
\(N\) & 4 & 16 & 36 & 64 & 168 & 288 & 440 & 624 & 1152 & 1760 & 2496 \\
\(n_1\) & 2 & 4 & 6 & 8 & 12 & 16 & 20 & 24 & 32 & 40 & 48 \\
\(n_2\) & 2 & 4 & 6 & 8 & 14 & 18 & 22 & 26 & 36 & 44 & 52 \\
\hline
\end{tabular}
\end{table}

\begin{flushleft}
\bf{S8. Additional experimental data for a multi-frequency signal detection}
\end{flushleft}

\begin{figure}[htbp]
\centering
\includegraphics[width=0.7\textwidth]{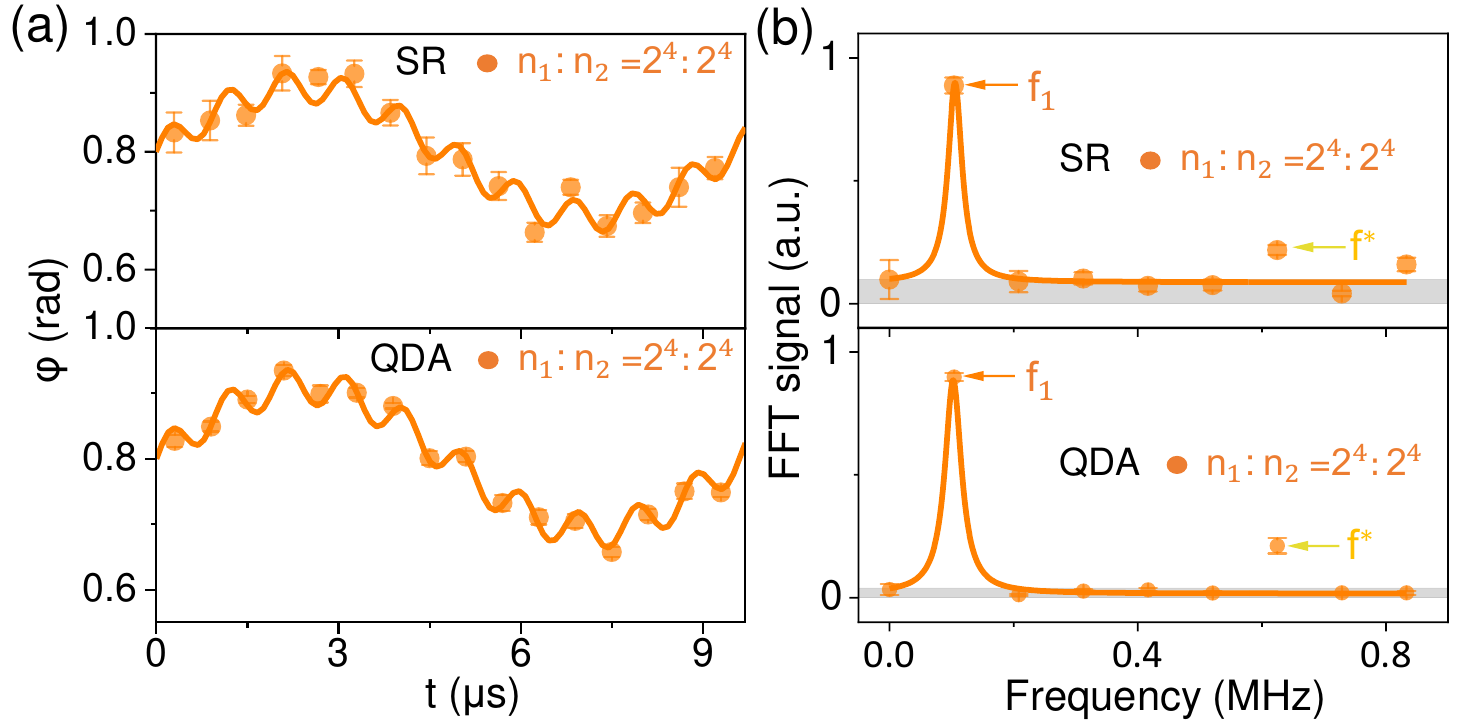}
\caption{Comparison of waveform reconstruction using SR and QDA methods within the ZOH reconstruction framework for a multi-frequency signal. (a) Results of multi-frequency waveform measurement with the SR and QDA method. The total number of resources is $N = {n_1}{n_2} = {2^8}$. The input waveform (solid line) and the recorded waveform data (orange dots) are given by ${b(t) = b[sin(2}\pi {f_1}t) + {\text{sin(2}}\pi {f_2}t)/4]$ with ${f_1} = 0.104$ MHz and ${f_2} = 1.04$ MHz. (b) The normal FFT signals of waveform measurement. The grey area indicates the measured noise level. When the sampling rate is small, the frequency aliasing is occurring and a false signal (${f^ * } = {f_s} - {f_2} = 0.625$ MHz) is detected in the experiment. We do not apply any window function for FFT processing.}
\label{fig:s5}
\end{figure}

For the same quantum resource for both cases as shown in Fig. 2(d), we further reduce the sampling rate to obtain a higher SNR measurement results for both cases as shown in Fig. \ref{fig:s5}. However, the sampling rate is lower than twice the highest frequency of the signal, it causes the original high-frequency signal to be sampled as a low-frequency signal ($f^*$) for both cases, which is the phenomenon of frequency aliasing. Hence, for a given total quantum resource count $N = {n_1}{n_2}$, the QDA method enables high-SNR, high-resolution spectral measurement of complex multi-frequency waveforms, even for weak high-frequency components that are unresolvable with the conventional SR method. This demonstrates that our method goes beyond single-parameter amplitude estimation for repeatable deterministic waveforms, and realizes the core waveform reconstruction function of a quantum oscilloscope within the ZOH reconstruction framework.

